\documentclass[preprint]{revtex4}

\usepackage{graphicx}
\usepackage{times}
\usepackage{epsfig}
\usepackage{amsmath}
\usepackage{amsfonts}
\usepackage{amssymb}
\usepackage{url}
\usepackage{hyperref}
\usepackage{subfigure}
\newcommand{\be}{\begin{equation}}
\newcommand{\ee}{\end{equation}}
\newcommand{\bea}{\begin{eqnarray}}
\newcommand{\eea}{\end{eqnarray}}

\def\ol#1{\overline{#1}}
\begin{document}

\pagestyle{plain}

\title{Flavor Physics in SO(10) GUTs with Suppressed Proton decay
Due to Gauged Discrete Symmetry}

\author{A. T. Azatov and R. N. Mohapatra }

\affiliation{Maryland Center for Fundamental Physics and
Department of
Physics,\\
 University of Maryland,\\ College Park, MD 20742, USA}

\date{February, 2008}

\preprint{\vbox{\hbox{UMD-PP-08-003}}}

\begin{abstract}
Generic $SO(10)$ GUT models suffer from the problem that Planck
scale induced non-renormalizable proton decay operators require
extreme suppression of their couplings to be compatible with
present experimental upper limits. One way to resolve this problem
is to supplement $SO(10)$ by simple gauged discrete symmetries
which can also simultaneously suppress the renormalizable R-parity
violating ones when they occur and make the theory ``more
natural''. Here we discuss the phenomenological viability of such
models. We first show that for both classes of models, e.g the
ones that use ${\bf 16}_H$ or ${\bf 126}_H$ to break B-L symmetry,
the minimal Higgs content which is sufficient for proton decay
suppression is inadequate for explaining fermion masses despite
the presence of all apparently needed couplings. We then present
an extended ${\bf 16}_H$ model, with three {\bf 10} and three {\bf
45}-Higgs, where is free of this problem. We propose this as a
realistic and ``natural'' model for fermion unification and
discuss the phenomenology of this model e.g. its predictions for
neutrino mixings and lepton flavor violation.
\end{abstract}
\maketitle 

\section{Introduction}
The neutrino observations of the past decade have put the
spotlight on gauged B-L symmetry as well as unification groups
such as SO(10) and $SU(2)_L\times SU(2)_R\times SU(4)_4$
containing B-L as prime candidates for theory of matter, forces
and flavor. While both these groups incorporate the seesaw
mechanism for neutrino masses, SO(10) has the additional
attractive feature that gauge couplings unify at high scale. It is
however highly nontrivial to obtain a ``truly natural'' SO(10)
model due to such issues as doublet triplet splitting, rapid
proton decay etc. In this paper we discuss how one aspect of this
naturalness can be addressed i.e. how one can naturally suppress
proton decay in SO(10) models while preserving our understanding
neutrino masses.

We first note that SO(10) models for neutrinos discussed in recent
literature can by and large be divided into two classes:

(i) One class which uses only renormalizable couplings involving
the Higgs fields {\bf 10}, {\bf 120} and {\bf 126} for fermion
masses and the last multiplet for breaking B-L symmetry and
multiplets such as {\bf 45} and/or {\bf 210} for gauge symmetry
breaking\cite{126}. This theory could be considered as an
ultraviolet complete theory by itself.

(ii)The  second class uses {\bf 10} plus ${\bf 16}\oplus {\bf
\bar{16}}$ for fermion masses with the {\bf 16}'s breaking the B-L
symmetry. Here one generally uses {\bf 45}+{\bf 54} Higgs fields
for SO(10) breaking. An important feature of this class is that it
has to rely on nonrenormalizable couplings to understand fermion
masses and therefore has to be viewed necessarily as an effective
theory at the GUT scale\cite{so1016}.

The first class of models leads to automatic R-parity conservation
when SO(10) breaks down to MSSM  so that there is a natural
candidate for dark matter whereas the second class of models
suffers from R-parity breaking and hence has no stable dark matter
in the absence of additional symmetries. So in principle one could
argue that this class of models are not ``pure'' SO(10) models.

 Both models have an additional naturalness problem arising from the fact
 that they allow R-parity conserving nonrenormalizable couplings
of the form $\lambda {\bf 16}^4_m/M_{Pl}$ which lead to rapid
proton decay. Such interactions could be induced by
nonperturbative Planck scale effects and it is therefore not safe
to ignore them. Present proton life time limits constrain
$\lambda$ to be $\leq 10^{-7}$. Such a small value of $\lambda$
would suggest that there is probably a symmetry responsible for
its smallness. This question is particularly urgent for the class
of SO(10) models with {\bf 16} Higgs since they rely on other such
dimension four higher dimensional operators with coefficients of
order one to understand fermion masses. This problem is generic to
all non-GUT susy theories such as MSSM or left-right models as
well as $SU(2)_L\times SU(2)_R\times SU(4)_4$ models and not just
GUT theories. One way to understand the suppression of such
operators despite the presence of non-perturbative gravitational
effects, is to have an additional gauge symmetry beyond SO(10)
which can forbid these unwanted terms. The simplest possibility is
to have a discrete gauge symmetry\cite{A1}. There are of course
other possibilities\cite{GUT}.

The discrete gauge symmetry supplemented SO(10) models that
suppress proton decay were studied for a large class of models in
a recent paper\cite{rmmr}. In particular two minimal SO(10)
models- one with  {\bf 16}-Higgs breaking the B-L symmetry and
another with {\bf 126} breaking B-L were shown to be free of both
proton decay problem as well as R-parity problem if  SO(10) was
supplemented by a gauged $Z_6$ symmetry. They looked promising for
phenomenology since all necessary terms in the superpotential for
phenomenology were allowed by the symmetry.
 It is the goal of this paper to study the viability of these models.

The results of this paper are the following: (i) the minimal
versions of both ${\bf 16}_H$-based as well as {\bf 126}-based
models discussed in Ref.\cite{rmmr} are not realistic since they
fail to give desired MSSM doublets that would be required to give
rise to realistic fermion masses and mixings; (ii) if the {\bf
16}-based models are extended to have three {\bf 10}-Higgs fields
and three {\bf 45} multiplets, one can have the desired
doublet-triplet splitting and fermion masses that can match
observations. This model differs from other {\bf 16}-based models
in that proton decay here arises only from the gauge boson
exchanges unlike other models where Planck scale induced effects
as well as Higgsino exchange ones play a role\cite{pdecay}; (iii)
we study the phenomenological implications of this model and
isolate some of its tests e.g. in the domain of lepton flavor
violation.

This paper is organized as follows: in sec. 2, we review the
salient features of the two classes of models; in sec. 3, we
discuss doublet-triplet splitting problem of the minimal models;
in sec.4 we discuss the three Higgs extension of the {\bf
16}-based model that fits fermion masses and mixings; in sec. 5,
we discuss how large neutrino mixings and observed neutrino masses
arise in this model. We summarize our results in sec. 6.

\section{The $SO(10)\times Z_6$ model for {\bf 16}-Higgs B-L breaking}

The main features of generic SO(10) models with {\bf 16}-Higgs
fields breaking B-L symmetry are the following: (i) the quarks and
leptons are assigned to three {\bf 16}-dimensional spinors
(denoted by $\psi_m$, m=1,2,3); (ii) the GUT symmetry is broken
down to $SU(2)_L\times SU(2)_R\times U(1)_{B-L}\times SU(3)_c$ by
a {\bf 45} $\oplus$ {\bf 54} set of Higgs fields; (iii)
$SU(2)_R\times U(1)_{B-L}$ symmetry is broken by the {\bf
16}-Higgs pair denoted by $\psi_H \oplus \bar{\psi}_H $ set to the
standard model symmetry which is then broken by $SU(2)$ doublets
that are linear combinations of the doublets in SO(10) $\bf{10}$ and
$\bf{16}$-Higgs
fields. The standard model symmetry along with supersymmetry
emerge just below the GUT sclae of  $2\times 10^{16}$ GeV.

 This model has several naturalness problems: it not only allows the
dangerous R-parity conserving $\frac{(16_m)^4}{M_Pl}$ terms but
also terms such as  $\frac{(16_m)^3 16_H}{M_Pl}$ terms which on
B-L symmetry breaking lead to all three types of R-parity
violating operators present in general MSSM i.e. $LLe^c$, $QLd^c$
and $u^cd^cd^c$ type. Thus this model for natural values of
couplings will lead to extremely rapid proton decay which is
unacceptable. The question addressed in Ref.\cite{rmmr} is to
search for
 gauged discrete symmetries that will keep the model
phenomenologically viable while keeping them ``proton decay safe''
and it was shown that the minimal anomaly free discrete gauge
symmetry
 is $Z_6$. Similar considerations for {\bf
126} type models also led to the symmetry $Z_6$ and in both cases
an extra  {\bf 10}-Higgs field denoted by $H'$ in addition to
those considered already.

To see the discrete symmetry charges for various fields that
forbid both R-parity violating terms as well as R-parity
conserving baryon number violating terms, while
 at the same time keeping the required terms responsible for good
phenomenology, we divide the superpotential terms into two
classes:
 type I terms that must be kept for phenomenology and type II
terms that must be forbidden to suppress proton decay and R-parity
violating terms. They are given below:

\noindent{\bf Terms of type I:} They include $\psi_m\,\psi_m\, H$,
$(\psi_m\,\overline{\psi}_H)^2/M_\mathrm{P}$,
$\psi_H\,\overline{\psi}_H$, $A^2$,  $S^{2,3}$ and $S\,A^2$, where
 $H$, $A$, $S$ are \textbf{10}-, \textbf{45}-,
\textbf{54}-plets, respectively.
Taking the discrete gauge symmetry to be $Z_N$, we can write down
the constraints on the $Z_N$ charges that are required by the type
I terms:
\begin{subequations}\label{C1}
\begin{align}
2q_{\psi_m}+q_{H} & = ~ 0\mod N\;, & \quad
2q_{\psi_m}+2q_{\overline{\psi}_H} & = ~ 0\mod N\;, \\
q_{\psi_H}+q_{\overline{\psi}_H} & = ~ 0\mod N\;,& \quad
q_{H}+q_{H'} & = ~ 0\mod N\;,\\
2q_A & = ~ 0\mod N\;,& \quad
2q_S & = ~ 0\mod N\;,\\
3q_S & = ~ 0\mod N\;,& \quad
2q_A+q_S & = ~ 0\mod N\;.
\end{align}
\end{subequations}
Here, we denote the $Z_{N}$ charge for a field $F$ by $q_F$.

\noindent{\bf Type II terms:} These are the terms that must be
forbidden from appearing in the superpotential and are
 $\psi_m\,\psi_m\,H'$, $\psi_m^4$,
$\psi_m\,\overline{\psi}_H$ and $\psi_m^3\,\psi_H$,
$\psi_m\,\psi_H\,H$, $\psi_m\,\psi_H\,H'$ and
$\psi_m\overline{\psi}_HA$. We forbid the $\psi_m\,\psi_m\,H'$ in
order to avoid large Higgsino mediated contribution to proton
decay since this is the very problem we are trying to solve. The
necessary constraints on the $Z_{N}$ charges have to be chosen
such that they satisfy the inequalities
\begin{widetext}
\begin{subequations}\label{C3}
\begin{align}
 2\,q_{\psi_m}+q_{H'} & \neq ~ 0\mod N\;,& \quad
 4\,q_{\psi_m}& \neq ~ 0\mod N\;, \\
 q_{\psi_m}+q_{\overline{\psi}_H} & \neq ~ 0\mod N\;,& \quad
 3\,q_{\psi_m}+q_{\psi_H}& \neq ~ 0\mod N\;,\\
q_{\psi_m}+q_{H',H}+q_{\psi_H} & \neq ~ 0\mod N\;,& \quad
q_{\psi_m}+q_{\overline{\psi}_H}+q_A & \neq ~ 0\mod N\;.
\end{align}
\end{subequations}
\end{widetext}

The last set of constraints come from the requirement that the
discrete symmetry must be a gauge symmetry i.e. it must be anomaly
free. The anomaly freedom constraints are:
\begin{widetext}
\begin{subequations}\label{C2}
\begin{eqnarray}
\quad 16\,(N_g\, q_{\psi_m}+q_{\psi_H}+q_{\overline{\psi}_H})
+10\,(q_{H}+q_{H'})+45\,q_{A}+54\,q_{S}&=&0\mod N'\;\\
2N_g\,q_{\psi_m}+2\,q_{\psi_H}+2\,q_{\overline{\psi}_H}+q_{H}+q_{H'}+
8q_{A}+12q_{S}&=&0\mod N\;\\
\text{where} ~N'=
\begin{cases}
N,~\text{odd}~N\\
N/2, ~\text{even}~N
\end{cases}
\end{eqnarray}
\end{subequations}
\end{widetext}
It was shown in \cite{rmmr} that the smallest symmetry allowing us
to fulfill all criteria is $Z_6$ for number of generation $N_g=3$.
A possible charge  assignments is $q_{\psi_m}=1$, $q_{\psi_H}=-2$,
$q_{\overline{\psi}_H}=+2$, $q_{H}=-2$, $q_{H'}=+2$, $q_{45,54}=0$
(cf.\ tables~\ref{tab:SO10} (a) and (b)). This charge  assignment
allows for seesaw couplings and the possibility of fermion masses
from couplings of type $\psi_m\,\psi_m\, H$. The allowed operator
$\psi_m\,\psi_m\, \overline{\psi}_H^2$ contributes to both  the
fermion masses as well as to the seesaw. The model also eliminates
the dangerous proton decay operator $Q\,Q\,Q\,L$ or operator of
type $(\psi_m)^4/M_\mathrm{P}$.

While the allowed set of operators provide a necessary condition
for the model being phenomenologically viable, the final step
where we judge whether it is acceptable first requires that we do
doublet triplet splitting and see if the sub-GUT scale structure
of the model can generate acceptable pattern of fermion masses or
not. We address this question in the next sub-section.

\subsection{Phenomenological Viability of the {\bf 16}-Higgs model}
To analyze the phenomenological implications of the model, let us
start by writing down the superpotential allowed by the discrete
symmetry and SO(10) invariance:
\begin{eqnarray}
W~&=&~W_Y~+~W_H\nonumber\\
 W_Y~&=&~h_i \psi_m\,\psi_m\,
H~+~\frac{\lambda_{1a}}{M_{Pl}}[\psi_m\,\psi_m\,
\overline{\psi}_H^2]_a~+~\frac{\lambda_2}{M_{Pl}}\psi_m\,\psi_m\,
A\,
H\nonumber\\
W_H&=&M H H'+ S H H' + A H H'+  \psi_H \psi_H
H+\bar\psi_H\bar\psi_H H'+M_{\psi}\bar\psi\psi
\end{eqnarray}
where $a$ denotes the various irreducible representations in the
product of two {\bf 16}'s.

\begin{table}[h]
\caption{} \label{tab:SO10}  \subtable[MSSM part]{\begin{tabular}{|c|c|}
 \hline
Field & quantum numbers\\
\hline
$\psi_m$ & $\bf{16}_1$\\
$H$ & $\bf{10}_{-2}$\\
$H'$ & $\bf{10}_{2}$\\
\hline
\end{tabular}}
\quad \subtable[$\bf{16}$-Higgs model.]
{\begin{tabular}{|c|c|}
\hline
Field & quantum numbers\\
\hline
$\psi_H$ & $\bf{16}_{-2}$\\
$\overline{\psi}_H$ & $\overline{\bf{16}}_{2}$\\
$A$ & $\bf{45}_{0}$\\
$S$ & $\bf{54}_{0}$\\
\hline
\end{tabular}}
\quad \subtable[$\bf{126}$-Higgs model.\!]{
\begin{tabular}{|c|c|} \hline
Field & quantum numbers\\
\hline
$\Delta$ & $\bf{126}_2$\\
$\ol\Delta$ & $\bf{126}_{-2}$\\
$\Sigma$ & $\bf{210}_{0}$\\
\hline
\end{tabular}}
\end{table}
 First we want to find whether we can solve doublet -triplet
splitting problem within this model field content. The vev s of
the {\bf 54} and {\bf 45} can be assumed to have the following
forms:
\begin{eqnarray}
<A> = \left ( \begin{smallmatrix}
& -1\\
1 & \\
\end{smallmatrix} \right )Diag(a,a,a, b,b)\nonumber\\
<S> = \left ( \begin{smallmatrix}
1& \\
 & 1\\
\end{smallmatrix} \right )Diag(s,s,s, -\frac{3}{2}s,-\frac{3}{2}s)\nonumber\\
\end{eqnarray}
It is useful to express all fields in terms of the SU(5)
multiplets.
\begin{eqnarray}
16=1+\ol 5+ 10\nonumber\\
\ol{ 16}=1+ 5 +\ol 10\nonumber\\
10=5+\ol 5
\end{eqnarray}

so the mass matrix in terms of SU(5) components of the fields will
look like
\begin{eqnarray}
(5_H,5_{H'},5_{\ol\psi_H})\left(\begin{array}{ccc} 0&M+A+S&
c\\M-A+S&0&0\\
0&c&M_\psi
\end{array}\right).\left( \begin{array}{c}
\ol{5}_{H}\\\ol{5}_{H'}\\\ol{5}_{\psi_H}\end{array}\right)
\end{eqnarray}
where $c$ is the vev of the $\bf 16$ and $\ol{16} $ of the
SO(10), $c=\langle \psi_H \rangle=\langle \bar\psi_H \rangle$.
and  to get expression for the mass matrix of the doublets(triplets)
one has to substitute instead of A(S) $~~b(-\frac{3s}{2})$ for doublets
and $a(s)$ for tripets respectively.
 One can see that this matrix can have a zero eigenvalue only if its
determinant vanishes i.e.
\begin{eqnarray}
\text{Det}=(M-A+S)\left( c^2-M_\psi(M+A+S)\right)=0;
\end{eqnarray}
This equation has two solutions: taking the first one i.e. $
M-A+S=0$, we find for the doublet mass matrix $M_{ud}$ in the
basis $(H,H',\psi_H)$ to be;
\begin{eqnarray}
M_{ud}=\left(
\begin{array}{lll}
 0 & z & c \\
 0 & 0 & 0 \\
 0 & c & M
\end{array}
\right)\nonumber\\
z=2b=2(M-\frac{3}{2}s).
\end{eqnarray}
To find the MSSM doublets in terms of the GUT submultiplets, we
diagonalize $M_{ud}$ and find its zero mode eigen-vector. The
usual MSSM Higgs fields $h_{u,d}$ will be linear combinations of
$(H,H',\psi_H)$ that correspond to the zero mode eigen-vector of
the above matrix. From the following equations, we find:
\begin{eqnarray}
\left(
\begin{array}{lll}
 0 & z & c \\
 0 & 0 & 0 \\
 0 & c & M
\end{array}
\right). \left(
\begin{array}{l}
 D_{11} \\
 D_{21}  \\
 D_{31}
\end{array}
\right)=0\nonumber\\
\left(
\begin{array}{lll}
 0 & 0 & 0 \\
 z & 0 & c \\
 c & 0 & M
\end{array}
\right). \left(
\begin{array}{l}
 U_{11} \\
 U_{21}  \\
 U_{31}
\end{array}
\right)=0\nonumber\\
\end{eqnarray}
It is easy to see that
\begin{eqnarray}
h_u~=~H'_u \\ \nonumber h_d~=~H_d
\end{eqnarray}
Since the {\bf 10}-Higgs denoted by $H'$ does not couple to {\bf 16}-fermions,
the MSSM up-Higgs doublet in this model does not couple to matter
and therefore all the up quarks remain massless. The solution (i)
to the determinant equation is therefore not acceptable.

Turning now to the second solution i.e. (ii)
$M+A+S=\frac{c^2}{M_\psi}$, we get the structure of the $M_{ud}$
mass matrix:
\begin{eqnarray}
M_{ud}=\left(
\begin{array}{lll}
 0 & \frac{c^2}{M} & c \\
 x & 0 & 0 \\
 0 & c & M
\end{array}
\right)
\end{eqnarray}
In this case the zero mode corresponding to the MSSM doublet
$h_d$, can be represented by the column vector:
\begin{eqnarray}
&D= \left(
\begin{array}{l}
0 \\
-\frac{M}{c\sqrt{1+\frac{M^2}{c^2}}} \\
 \frac{1}{\sqrt{1+\frac{M^2}{c^2}}}
\end{array}
\right) \nonumber\\
&U = \left(
\begin{array}{l}
 -\frac{M}{c\sqrt{1+\frac{M^2}{c^2}}} \\
 0  \\
  \frac{1}{\sqrt{1+\frac{M^2}{c^2}}}
\end{array}
\right)
\end{eqnarray}
The MSSM doublet $h_d$ in this case does not couple to quarks and
charged leptons and leave those fields massless. Again this is not
acceptable. Taking these two cases together we conclude that in
the minimal gauge discrete symmetric $Z_6$ model, the doublet
triplet splitting and nontrivial fermion masses cannot happen
simultaneously and the model is therefore not phenomenologically
viable. We wish to emphasize again that in drawing this conclusion, we
have also considered higher dimensional operators that could contribute
fermion masses. It turns out that in our, case operators such as
 $(\psi_m)^2(\psi_H)^2$, $(\psi_m)^2 H' A$ which can lead to the bottom
quark  mass are not allowed due to the $Z_6$ charge assignments.

\subsection{Phenomenological viability of the $SO(10)\times Z_6$
$\textbf{126}$ model} This class of models typically have the
Higgs multiplets of type {\bf 10} {\bf 126},{\bf 210} (and {\bf
120}) fields to explain fermion masses including neutrino masses
and mixings. These models do not have R-parity breaking terms even
after GUT symmetry breaking. The $Z_6$ charge assignments that
makes the model proton-decay-safe while keeping necessary terms
for possible fermion masse are given in Table \ref{tab:SO10} (c).

The Higgs superpotential in this case looks like

\begin{eqnarray}
W=M_H H H' +M_{\Delta} \Delta \ol \Delta +\lambda_{\Delta \Sigma
H}\Delta \Sigma H +M_\Sigma \Sigma^2 + \lambda_{\Sigma^3}\Sigma^3
+\lambda_{\ol\Delta \Sigma H'}\ol\Delta \Sigma H' +\lambda_{\ol
\Delta \Sigma \Delta}\ol \Delta \Sigma \Delta.
\end{eqnarray}

One might think that this can lead to a realistic model for
fermion masses. However as in the previous sub-section, we must
analyze the doublet-triplet splitting in order to study the
fermion masses. we will see that in this case too there is a
conflict between the doublet-triplet splitting and fermion masses.

To see this, we write down the mass matrix  for MSSM doublets
contained in various GUT Higgs multiplets in the theory: (see
\cite{Fukuyama} for the exact Clebsch -Gordon coefficients)
\begin{eqnarray}
&(H_u,H'_u,\Delta_u,\ol \Delta_u,\Sigma_u)\times M_{ud}\times
\left(\begin{matrix}
H_d\\
H'_d\\
\ol\Delta_d\\
\Delta_d\\
\Sigma_d
\end{matrix}
\right);\nonumber\\
&M_{ud}=\left (
\begin{smallmatrix}
 0 &M_H &0 &-\frac{\lambda_{H\Delta
\Sigma}}{\sqrt{10}}(\Phi_2+\frac{\Phi_3}{\sqrt{2}})&
-\frac{\lambda_{H\Delta \Sigma} }{\sqrt{5}} v\\
 M_H &0 &\frac{\lambda_{H' \ol\Delta \Sigma}}{\sqrt{10}}(\Phi_2-
\frac{\Phi_3}{\sqrt{2}}) &0&0\\
  \frac{\lambda_{H\Delta \Sigma}}{\sqrt{10}}(\Phi_2-\frac{\Phi_3}{\sqrt{2}}) &0
  & M_\Delta+\frac{\lambda_{\Sigma \Delta\ol\Delta}}{15\sqrt{2}}(\Phi_2
-\frac{\Phi_3}{\sqrt{2}})&0&0\\
0&-\frac{\lambda_{H'\ol\Delta\Sigma}}{\sqrt{10}}(\Phi_2+\frac{\Phi_3}{\sqrt{2}})
& 0&M_\Delta+\frac{\lambda_{\Sigma
\Delta\ol\Delta}}{15\sqrt{2}}(\Phi_2+\frac{\Phi_3}{\sqrt{2}})&
\frac{\lambda_{\ol\Delta\Delta \Sigma} }{10} v\\
0&-\frac{\lambda_{H'\ol\Delta \Sigma} }{\sqrt{5}}\ol
v&0&\frac{\lambda_{\ol\Delta\Delta \Sigma} }{10} \ol v&M_\Sigma
+\frac{\lambda_{\Sigma^3}}{\sqrt{2}}(\Phi_2+\frac{\Phi_3}{\sqrt{2}})
\end{smallmatrix}
\right)
\end{eqnarray}
Where $\Phi_1,\Phi_2,\Phi_3$ are the vevs of $\Sigma$ in different directions and $v,\ol v$  vevs of $\Delta,\ol\Delta$ fields respectively. For simplicity of analysis, we rewrite this matrix in the symbolic
form as follows:
\begin{eqnarray}
&M_{ud}=\left(
\begin{array}{lllll}
 0 &m &0 & \frac{d a}{b}&c\\
 m &0 &b &0&0\\
  a &0& M&0&0\\
0&d&
0&M_2&A\\
0&c_1&0&A&B
\end{array}
\right)
\end{eqnarray}
We want to have one massless state so we require that
\begin{eqnarray}
Det M_{ud}=0.
\end{eqnarray}
 This requires that one of the following two conditions be
satisfied $a=\frac{m M}{b}$ or
 $a=\frac{A b c_1  d-A^2 b m -b c c_1 M_2 +B b m M_2 }{d (B d-A c_1)}$.
Let us now investigate the first one (i)$ a=\frac{m M}{b}$ which
leads to the following zero mass eigenstate:
\begin{eqnarray}
U=\left(
\begin{array}{l}
 0 \\
 -\frac{M}{b\sqrt{1+\frac{M^2}{b^2}}} \\
  \frac{1}{\sqrt{1+\frac{M^2}{b^2}}} \\
0\\
0
\end{array}
\right)
\end{eqnarray}
\begin{eqnarray}
D=\left(
\begin{array}{l}
 -\frac{b}{m\sqrt{1+\frac{b^2}{m^2}}} \\
 0 \\
  \frac{1}{\sqrt{1+\frac{b^2}{m^2}}} \\
0\\
0
\end{array}
\right)
\end{eqnarray}
In this case we see that the up-MSSM Higgs doublet does not couple
to matter fermions.

Turning now to the case (ii) where we have $a=\frac{A b c_1  d-A^2 b m -b c c_1 M_2 +B b m M_2 }{d(B d-A c_1)}$, we get for the same
eigenstates:
\begin{eqnarray}
U=\left(
\begin{array}{l}
 * \\
 0  \\
  0 \\
*\\
*
\end{array}
\right)
\end{eqnarray}
\begin{eqnarray}
D=\left(
\begin{array}{lllll}
 0 \\
 * \\
  0 \\
*\\
*
\end{array}
\right)
\end{eqnarray}
where we have written  only the zero entries in the columns. The
$*$'s represent non-zero entries whose detailed form is irrelevant
for our discussion. It is clear that in both cases the doublet
-triplet splitting is incompatible with giving masses to the
fermions; in the case (i) to up quarks and in case (ii) to the
down quarks.

So neither the minimal {\bf 16} nor {\bf 126} models when made
proton decay safe can lead to viable fermion masses along with
doublet triplet splitting. We therefore have to extend the Higgs
sector to get a realistic model. In the next section, we give one
such example and analyze its flavor phenomenology.

\section{Extended 16-Higgs model}
We now extend the {\bf 16 }-Higgs model by adding extra Higgs
multiplets in such a way that anomaly freedom as well as proton
decay constraints are satisfied and yet the model can lead to
viable phenomenology. The simplest possibility appears to be to
extend our $\textbf{16}$ model by adding one additional
$\textbf{10}$ - $H_3$ and two $\textbf{45}$ - ${A_2}$, ${A_3}$
fields under SO(10), where $H_3$ has zero charge under $Z_6$, and
$A_{2,3}$ charges are -2 and 2 respectively (see table below).
\begin{table}[h]
\caption{} {
\begin{tabular}{|c|c|} \hline
Field & quantum numbers\\
\hline
$A_2$ & $\bf{45}_{-2}$\\
$A_3$ & $\bf{45}_{2}$\\
$H_3$ & $\bf{10}_{0}$\\
\hline
\end{tabular}}
\end{table}
$Z_6$ charges are easily seen not to ruin our anomaly cancellation
conditions. We also have redefined our $H', H$ fields as $H_{1,2}$
and $A$ as $A_1$ for the simplicity of notation . Now the
superpotential is given by:
\begin{eqnarray}
W=&M H_1 H_2+S H_1 H_2 +A_1 H_1 H_2+\psi_H \psi_H
H_1+\bar\psi_H\bar\psi_H H_2+M_{\psi}\bar\psi_H\psi_H+ M_3
H_3^2+\nonumber\\
 &+H_2
H_3 A_2+H_1 H_3 A_3 +\frac{H_3 A_3 \ol \psi_H
\ol\psi_H}{M_{Pl}}+\frac{H_3 A_2 \psi_H \psi_H}{M_{Pl}}
\end{eqnarray}
Our model allows an operator of the form $\frac{\psi_m^4
A_3}{M_{Pl}}$ where substituting the vev of the field $A_3$ we get a
proton decay operator with effective $\lambda \simeq
\frac{M_U}{M_{Pl}}\ll 1$ but not suppressed enough to be acceptable.
However this problem disappears if the vev $\langle A_3\rangle
=0$. We will see below that there is an allowed vacuum, where
indeed this is possible.

To study the doublet triplet splitting in this model, note that
the mass matrix for the $\textbf{5}$  and $  \ol{\bf 5}$ of SU(5)
is given by
\begin{eqnarray}
\label{16ext} \left(\begin{array}{cccc}
0&M+A_1+S&0&c\\
M-A_1+S&0&A_2&0\\
0&-A_2&M_3&0\\
0&c&\delta &M_\psi
\end{array}\right)_{ud}
\end{eqnarray}
Where $\delta$ in the (43) element of the matrix comes from $\frac{H_3 A_2 \psi_H \psi_H}{M_{Pl}}$ coupling.
As before, we want the determinant of this matrix to vanish. This
leads to the following constraints

\noindent Case (i)
\begin{eqnarray} M-A_1+S=0
\end{eqnarray}
\noindent Case(ii):
\begin{eqnarray}
M+A_1+S=\frac{c^2 M_3 +c A_2 \delta}{M_3 M_\psi}
\end{eqnarray}
Here we consider only the simpler of the two cases above i.e. case
(i) to illustrate that our proposal leads to a realistic model. In
the first case $M-A_1+S=0$ , $D=(1, 0, 0, 0)$ (implying that the
$h_d$ has non-zero component in the multiplet $H_1$) in the same
way as was in the minimal model, but now due to the presence of
$A_2$ field all the $U_{1}$ is nonvanishing, so that the "up"
quarks will get masses from the $\psi_m^2 H_1$ operator. In the
next sections we will discuss the detailed fit to fermion masses
for this extended $\textbf{16}$ model.

As we can see extended \textbf{16} model can solve doublet-triplet
splitting problem as well as provide masses for all fermions, but
now we have to check whether higgsino mediated proton decay
operators are allowed. Even though quarks and leptons  couple only
to the $H_1$ field and there is no mass term $ \propto H_1 H_1$, mixing
between $H_1,H_2,H_3,\psi_H$ fields can lead to the nonvanishing
diagrams with higgsino exchange. The contribution of these
diagrams will vanish if only the $(H_1 H_1)$ element of the
inverse mass matrix (\ref{16ext}) for the heavy triplets  vanishes, thus the
triplet part of the $A_2$ should be zero. We will see in the next
section that the requirement of the $<A_3>=0$ combined with F
flatness condition will lead to this condition.

\section{Supersymmetry down to the weak scale}
First we want to find out whether there is a minimum of the
potential that can correspond to the solution we are interested in
i.e. having supersymmetry survive down to the weak scale. The
Higgs part of the superpotential is:
\begin{eqnarray}
W_H=M_\psi \psi_H\ol\psi_H+m_1 A_1^2+m_2 A_2 A_3+m_s S^2+\lambda_1
A_1^2 S+\lambda_2 S^3+\lambda_3 A_2 A_3 S +\lambda_4 \ol \psi_H
\psi_H A_1+{\lambda}_5 A_1 A_2 A_3
\end{eqnarray}
+nonrenormalizable terms

The vev of the \textbf{45}, \textbf{54} and \textbf{16} fields
will in general have the following form:
\begin{eqnarray}
\langle A_i \rangle=\left ( \begin{smallmatrix} & 1\\
-1&
\end{smallmatrix}\right )(a_i,a_i,a_i,b_i,b_i);\nonumber\\
\langle S \rangle=\left ( \begin{smallmatrix} 1& \\
&1
\end{smallmatrix}\right )(s,s,s,-\frac{3}{2}s,-\frac{3}{2}s);\nonumber\\
\langle \ol \psi_H\rangle =\langle\psi_H\rangle=c
\end{eqnarray}
So we can rewrite the superpotential in terms of the vev of these
fields, using the further identities:
\begin{eqnarray}
M_\psi \psi_H\ol\psi_H&=&M_\psi c^2\nonumber\\
m_1A_1^2&=&-2(3a_1^2+2 b_1^2)m_1\nonumber\\
m_2A_3A_2&=&-2(3a_2a_3+2 b_2 b_3)m_2\nonumber\\
\lambda_1A_1^2 S&=&(-6a_1^2 s+6 b_1^2 s)\lambda_1\nonumber\\
\lambda_3 A_2 A_3 S&=&(-6a_2 a_3 s+6 b_2 b_3)\lambda_3 s\nonumber\\
\lambda_2 S^3&=&-\frac{15}{2}\lambda_2 s^3\nonumber\\
m_sS^2&=&15 m_s s^2
\end{eqnarray}
The condition of the vanishing F terms leads to the following
constraints,
\begin{eqnarray}
\label{fflattness} \frac{\partial W}{\partial a_1}&=&-12 m_1 a_1
-12\lambda_1 a_1 s+3
\lambda_4 c^2=0\nonumber\\
\frac{\partial W}{\partial b_1}&=&-8 m_1 b_1 +12\lambda_1 b_1 s+2
\lambda_4 c^2=0\nonumber\\
\frac{\partial W}{\partial a_2}&=&-6 m_2 a_3 -6\lambda_3 a_3 s=0\nonumber\\
\frac{\partial W}{\partial b_2}&=&-4 m_2 b_3 +6\lambda_3 b_3 s=0\nonumber\\
\frac{\partial W}{\partial a_3}&=&-6 m_2 a_2 -6\lambda_3 a_2 s=0\nonumber\\
\frac{\partial W}{\partial b_3}&=&-4 m_2 b_2 +6\lambda_3 b_2 s=0\nonumber\\
\frac{\partial W}{\partial s}&=&30m_s s+\lambda_1(6 b_1^2-6a_1^2)+\lambda_3(-6a_2 a_3+6 b_2 b_3)-\frac{45}{2}\lambda_2 s^2=0\nonumber\\
\frac{\partial W}{\partial c}&=&2c M_\psi+2 c \lambda_4(3 a_1+2
b_1)=0
\end{eqnarray}
we are interested in whether there exist a solution with
$a_3=b_3=0$ and $b_2\neq 0$ these constraints lead to the
following restrictions on the vevs
\begin{eqnarray}
\label{fflattness2} a_2=0,~~
s&=&\frac{2m_2}{3\lambda_3}\nonumber\\
(3a_1+2 b_1)\lambda_4&=&-M_\psi,~~s=\frac{2
m_1(b_1-a_1)}{\lambda_1(2 a_1+3 b_1 )}
\end{eqnarray}
required to suppress higgsino exchange diagrams. Now we will
present the other massless components of the higgs fields that
provide the breaking of the $SO(10)\rightarrow SU(3)\times
SU(2)\times U(1)$. We will identify them by their charges under
$SU(3)\times SU(2)\times U(1)$

1) (3,1,2/3) fields $( A_1,A_2,A_3,\psi_H)$;

\begin{eqnarray}
\left(\begin{matrix} -4 m_1-4\lambda_1 s &0&0&-2 \lambda_4
c\\
0&0&-2m_2 -2\lambda_3 s +2 i a_1 \lambda_5 &0\\
0&-2m_2 -2\lambda_3 s -2 i a_1 \lambda_5&0&0\\
-2 \lambda_4 c &0&0 &\lambda_4(2b_1-a_1)+M_\psi
\end{matrix}
\right)
\end{eqnarray}
2) (3,2,-5/6) fields $( A_1,A_2,A_3,S)$ ;
\begin{eqnarray}
\left(\begin{matrix} -4 m_1+\lambda_1 s &0&-i
b_2\lambda_5&2i(a_1+b_1)\lambda_1\\
0&0&-2m_2 -\lambda_3 \frac{s}{2} +i( a_1-b_1) \lambda_5 &0\\
i b_2 \lambda_5&-2m_2 -\lambda_3 \frac{s}{2} -i( a_1-b_1) \lambda_5 &0&i\lambda_3 b_2\\
-2i(a_1+b_1)\lambda_1 &0&-i\lambda_3 b_2 &4 m_s-3\lambda_2 s
\end{matrix}
\right)
\end{eqnarray}

3) (3,2,1/6)  fields $ ( A_1,A_2,A_3,S,\psi_H)$;

\begin{eqnarray}
\left(\begin{smallmatrix}
 -4 m_1+\lambda_1 s&    0 & -ib_2\lambda_5&   2i(a_1-b_1)\lambda_1&   -2\lambda_4 c\\
0                   &   0&-2m_2+\lambda_3\frac{s}{2} +i( a_1+b_1) \lambda_5 &0   &   0               \\
i b_2 \lambda_5     &-2m_2+\lambda_3\frac{s}{2} -i( a_1+b_1) \lambda_5 &0&  -ib_2\lambda_3       &0\\
-2i(a_1-b_1)\lambda_1 &0&ib_2\lambda_3 &4m_s-3\lambda_2s&0\\
-2\lambda_4 c&0&0&0&M_\psi+\lambda_4 a_1
\end{smallmatrix}
\right)
\end{eqnarray}
4) (1,1,1) fields $( A_1,A_2,A_3,\psi_H)$;

\begin{eqnarray}
\left(\begin{matrix} -4 m_1+6\lambda_1 s &0&-2 i \lambda_5 b_2&2
\lambda_4
c\\
0&0&-2m_2 +3\lambda_3 s +2 i b_1 \lambda_5 &0\\
2 i \lambda_5 b_2&-2m_2 +3\lambda_3 s +2 i b_1 \lambda_5 &0&0\\
2 \lambda_4 c &0&0 &\lambda_4(3a_1-2 b_1)+M_\psi
\end{matrix}
\right)
\end{eqnarray}
from the equations (\ref{fflattness}-\ref{fflattness2}) one can
see that each of these matrices will have one massless eigenstate.
So we have total 32 massless goldstone bosons. One more goldstone
boson needed to break SO(10) down to SU(2)X SU(3)X U(1) comes from
the phase of the $\psi_H,\ol\psi_H$ fields

\section{Fermion masses}
 The following couplings allowed by $Z_6\times SO(10)$
 symmetries will lead to fermion masses after symmetry breaking.
\begin{eqnarray}
W=h^{10}\psi_m\psi _m H_1+
f^{10}\frac{\psi_m^2\ol\psi_H^2}{M_{Pl}}+f^{126}\frac{\psi_m^2\ol{\psi}_H^2}{5!
M_{Pl}}+k^{120}\frac{\psi_m^2AH_1}{3!
M_{Pl}}+g^{120}\frac{\psi_m^2A^2H_1}{3!M_{Pl}^2}
+g^{126}\frac{\psi_m^2A^2H_1}{5!M_{Pl}^2}
\end{eqnarray}

Where $h^{10},f^{10},f^{126},g^{126}$ are symmetric $3\times 3$
and $k^{120}, g^{120}$ antisymmetric matrices. The upper index of
$h^{10},f^{126}...$, $\textbf{10},\textbf{120},\textbf{126}$ shows
the SO(10) structure of the fermion couplings.
\begin{eqnarray}
M_u&=&v_u\left[\alpha\left( h^{10}-k^{120}\frac{a-b}{M_{Pl}}+
g^{120}\frac{b(a-b)}{M_{pl}^2} +g^{126}\frac{ab+a^2}{M_{Pl}^2}
\right)+\xi\left( f^{10}\frac{4c}{M_{Pl}}
-f^{126}\frac{8c}{M_{Pl}}\right)
\right]\nonumber\\
M_\nu^D&=&v_u\left[\alpha\left( h^{10}+k^{120}\frac{3a+b}{M_{Pl}}-
g^{120}\frac{b(3a+b)}{M_{pl}^2}
-g^{126}\frac{3(ab+a^2)}{M_{pl}^2}\right)+\xi\left(
f^{10}\frac{4c}{M_{Pl}} +f^{126}\frac{24c}{M_{Pl}}\right)
\right]\nonumber\\
M_d&=&v_d \gamma\left[ h^{10} -k^{120}\frac{a+b}{M_{Pl}}
-g^{120}\frac{b(a+b)}{M_{pl}^2}+g^{126}\frac{-ab+a^2}{M_{pl}^2}\right]
\nonumber\\
M_e&=&v_d \gamma\left[ h^{10} +k^{120}\frac{3a-b}{M_{Pl}}
+g^{120}\frac{b(3a-b)}{M_{pl}^2}-g^{126}\frac{3(-ab+a^2)}{M_{pl}^2}\right]
\nonumber\\
 M_\nu^M&=&16f^{126}\frac{(\xi v_u)^2}{M_{pl}}-(M_\nu^D)^T(16 f^{126}
\frac{c^2}{M_{Pl}})^{-1}M_{\nu}^D\nonumber\\
\end{eqnarray}

Where the vev of the fields $H^1$ and $\ol\psi_H$ are related to
the vev of the MSSM doublets $h_u$ and $h_d$ in the following way
 \bea \langle
H_{1u}\rangle =\alpha \langle h_u
\rangle\nonumber\\
\langle H_{1d}\rangle =\gamma \langle h_d
\rangle\nonumber\\
\langle {\ol{\psi}_H }_u\rangle =\xi \langle h_u \rangle
 \eea

 Our claim is that these Yukawa coupling structure is rich
enough to fit all the fermion masses. We give below an example of
a scenario where correct fermion masses can arise.

We take the case where all antisymmetric couplings vanish and that
down quarks and leptons are brought to the diagonal basis at the
same time. Thus $ g^{126}$ and $h^{10}$ are diagonal, and all the
mixing in the quark and lepton sector  arise from the couplings
$f^{10}$ and $f^{126}$. We now show that even under such limiting
assumptions we can fit all the fermion masses. We know the quark
masses at the GUT scale thus we can find corresponding
$h^{10},g^{126}$, but on the other hand in the case when all the
quark mass matrices are symmetric the mass matrix for the up
quarks is equal to
\begin{eqnarray}
M_u=(CKM)^T.M_u^{diag}.CKM;
\end{eqnarray}
this leads to the constraint on the linear combination of $f^{10}$
and $f^{126}$.On the other hand we can find $f_{126}$ from the
known neutrino masses and mixing angles \cite{nfit}, so this fixes
$f^{10}$ and $f^{126}$.

\begin{figure}[h]
\includegraphics[scale=1,angle=0]{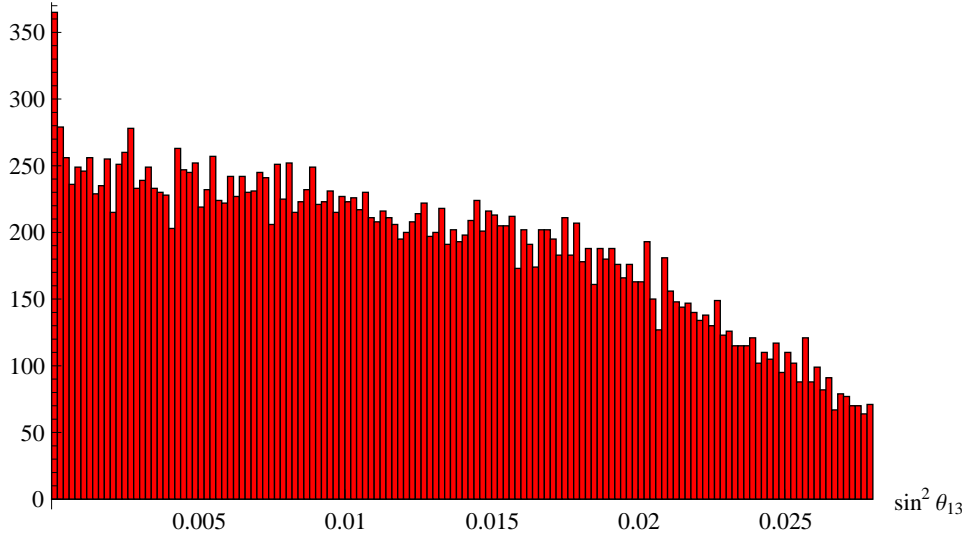} \label{figferm1}
\caption{\label{hist} Distribution plot for the values of $sin^2
\theta_{13}$}
\end{figure}

So here is the fit for $a,b,c,\alpha,\xi$ that leads to the good
quark and lepton mixing. We set $tan(\beta)=55$,  the masses of
the quarks and leptons and the vev's $v_u$, $v_d$ at the GUT could
be found in \cite{Das}.
\begin{eqnarray}
\alpha &=&0.8\nonumber\\
\xi&=&-0.47\nonumber\\
\gamma &=& 1\nonumber\\
a &=&0.028 M_Pl \nonumber\\
b &=&-0.014 M_Pl\nonumber\\
c &=&-0.024 M_Pl \nonumber\\
\end{eqnarray}

\begin{eqnarray}
h_{10}&=&Diag(0.000574402,0.0208157,0.792778)\nonumber\\
g_{126}&=&Diag(0.114692, -4.43859, 0.190027)\nonumber\\
 f_{10}&=&\left (
 \begin{matrix}
-0.00858937 - 0.000536445 i & -0.0110172 +
      0.00138491 i&
    0.094032 - 0.039795 i\\
    -0.0110172 +
      0.00138491 & -0.286505 +
      0.000708062 i& -0.508733 -
      0.00956549 i\\
       0.094032-
      0.039795 i& -0.508733 -
      0.00956549 i& -0.134644 - 0.0176603 i
 \end{matrix}\right)\nonumber\\
 f_{126}&=&\left (\begin{matrix}
-5.509\times 10^{-6}+5.992 \times 10^{-6} i &0.00003070
-0.00001463 i&-0.001095+0.0004482 i\\
0.00003070
-0.00001463 i&-0.00014401+9.1768\times 10^{-6}i&0.004929-0.00003607 i\\
-0.001095+0.0004482 i&0.004929-0.00003607 i &-0.1718 - 0.008830 i
\end{matrix}\right )
\end{eqnarray}
these Yukawa couplings lead to the good mass matrices for the up,
down quarks, charged leptons and neutrinos. Note that there are no
dimension five operators that contribute to down quark mass matrix
due to the discrete symmetry of the model.
\begin{eqnarray}
M_d&=&Diag(1.46323, 32.2949, 1638.17)  \text{MeV}\nonumber\\
M_e&=&Diag(0.35668, 75. ,1636.) \text{MeV}\nonumber\\
M_u&=&Diag(0.757795, 208.466,87029.8)\text{MeV}\nonumber\\
M_u&=&\left( \begin{matrix}
14.74-3.341 i&-67.49 + 8.615 i& 586.2-247.9 i\\
-67.49 + 8.615& 312.8 + 4.202i&-3159. -57.83i\\
 586.2-247.9 i&-3159. -57.83i& 86910.
 \end{matrix}\right ) \text{MeV}\nonumber\\
v_u &=&135016.\text{MeV},~~v_d =2065.81 \text{MeV}\nonumber\\
M^2_\nu&=&Diag(9.577*10^{-7},0.00007594,0.002697) \text{eV}^2\nonumber\\
U_{ei}&=&\left( \begin{matrix}
-0.7995&0.59481&-0.083258\\
-0.35036 - 0.14031 i&-0.51466 - 0.090935 i&-0.31226 + 0.69779
i\\
-0.46449 - 0.050427 i&-0.59179 - 0.15112 i&0.2327 - 0.5954 i
 \end{matrix}\right )\nonumber\\
V_{CKM}&=&\left( \begin{matrix}0.973841&
    0.227198&
    0.00169092-0.00292876 i\\
    -0.227079-0.000134603 i&
    0.97298-0.000031403i&
    0.0369876\\
    0.00675874-0.00284968 i&-0.0364044-0.000664834i&
    0.99912
\end{matrix}\right )
\end{eqnarray}
The FIG.\ref{hist} shows the distribution for the values of
$sin^2\theta_{13}$ in our model for $2\sigma$ values of neutrino
masses and mixings. As is clear from this figure the model has a
slight preference towards the region of small $~\theta_{13}$.  The
distribution for the other parameters of the neutrino mass
matrices appear to be spread uniformly over the allowed regions.

\section{Constraints from the lepton flavor violation}
In this section, we discuss the predictions of this model for
lepton flavor violation. As is well
known\cite{Hisano,masiero:2007ha}, even if the slepton mass
matrices are diagonal at the GUT scale the RGE running down to the
scale of the righthanded neutrino will lead to the mixing in the
slepton sector, which via one loop diagrams leads to lepton
violation. We will assume mSUGRA boundary condition for scalar
partner masses and use the renormalization group equations to run
them down to the seesaw scale when the right handed neutrinos
decouple.

 We will work in the
basis with diagonal righthanded majorana neutrino matrix, then the
slepton mixing will be approximately equal to
\begin{eqnarray}
(\delta_{ij}^l)_{LL}=-\frac{3 m_0^2 + A_0^2}{8\pi^2 m_0^2}
\sum_{k=1}^3(Y_\nu)_{ik}(Y^*_\nu)_{jk}\ln(\frac{M_{GUT}}{M_{R_k}})
\end{eqnarray}
where $Y_\nu$ are the Yukawa couplings of the Dirac neutrino.
These Yukawa couplings appear to be of roughly
\begin{eqnarray}
Y_\nu\sim \left (\begin{matrix}
10^{-5}&5 \cdot10^{-4}& 5\cdot 10^{-3}\\
10^{-5}&5 \cdot10^{-3}&2\cdot 10^{-2}\\
 10^{-5}&3 \cdot10^{-3}&0.3
\end{matrix}
\right )
\end{eqnarray}
Here $Y_\nu$ is a linear combination of the Yukawa couplings
$h_{10}, f_{10}$ and $f_{126}$ of the previous section. The
slepton mixing leads to the lepton flavor violating processes
$l_i\rightarrow l_j \gamma$ with the amplitude equal to \bea iM=e
m_{li}\epsilon^\lambda \ol l_j\left(i q^\mu \sigma_{\lambda\mu}
(A_L P_L+A_R P_R)\right) l_i \eea Where the $q$ is the momentum of
the photon and $P_{L,R}=\frac{1}{2}(1\mp\gamma_5)$, the exact
expression for the $A_{L,R}$ can be found in
\cite{masiero:2007ha}. The branching ratio for this processes will
be equal to 

\begin{eqnarray}
 &\frac{BR(l_i\rightarrow l_j \gamma)}{BR(l_i\rightarrow l_j
\nu_i \ol \nu_j)} =\frac{48 \pi^3
\alpha}{G_F^2}(|A_L^{ij}|^2)\nonumber\\
&A_l^{ij}\propto \frac{\alpha_2}{4\pi}\frac{(\delta_{ij}^l)_{LL}}{\tilde{m}_{0}^2}
\end{eqnarray}


The present bounds on this processes are\cite{meg}

\begin{table}[h]
{
\begin{tabular}{|c|c|} \hline
$BR(\mu\rightarrow e \gamma)$ & $ 1.2\cdot 10^{-11}$\\
\hline
$BR (\tau\rightarrow \mu \gamma)$&$ 6.8\cdot 10^{-8}$\\
\hline
\end{tabular}
}
\end{table}
We will carry out our calculations for the branching ratio in the
mSUGRA scenario, where there are only four parameters that will
fix the low energy values of the slepton masses $M_{1/2},m_0,A_0,
tan\beta, sign (\mu)$ but our fit for the fermion masses was
carried out for the $tan\beta=55$ so we will stay with this value.
\begin{figure}[h]
\includegraphics[scale=1,angle=0]{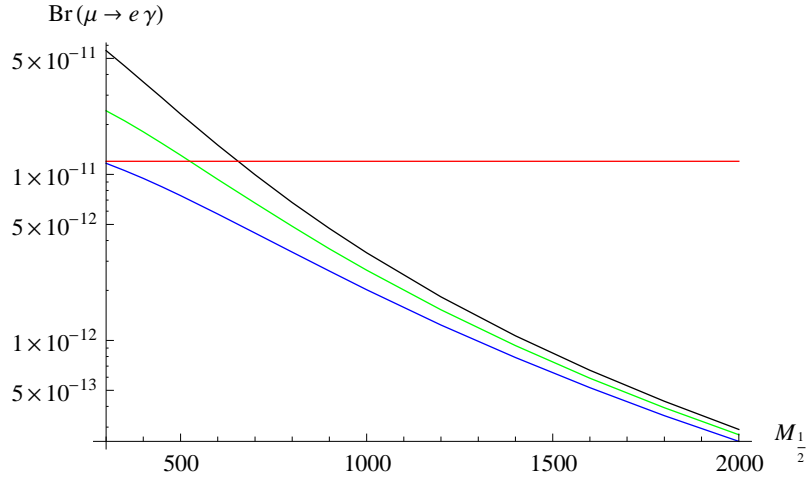} \label{figferm1}
\caption{\label{lfv}Branching ratios $BR(\mu\rightarrow e
\gamma)$ for different values of $m_0$ (black 300 Gev, green 400
Gev, blue 500 Gev), $A_0=0, tan\beta=55$, sign($\mu$)=1}
\end{figure}
In the FIG.\ref{lfv} one can see dependence of the branching
ratios on the $M_{1/2}$ for the fixed values of $m_0,
tan\beta,A_0,sign(\mu)$. We note that branching ratio for $\mu\to
e+\gamma$ for almost the entire parameter range of our model is
above $10^{-13}$ a value which is in the accessible range of the
ongoing MEG experiment\cite{meg}.

\section{Comments}
We add a few comments on the model described before closing:

(i) In this model, the leading order proton decay operator is
$\frac{\psi^4_m A^2_2}{M^3_{Pl}}$. After GUT symmetry breaking
this leads to the effective strength $\lambda \sim
\frac{M^2_U}{M^2_{Pl}}$. Naively this is of order $2\times
10^{-5}$, bigger than the present upper limit but is a
considerable improvement in the naturalness. It could also be that
the GUT vev could arise mainly from $A_1$ with $\langle
A_2\rangle$ being an order of magnitude smaller. This would then
give the desired suppression to proton decay. In that case this
will be the dominant graph for proton decay. Note that there are
no Higgsino mediated diagrams for proton decay in this model. In
addition, there is the gauge exchange diagram, present in all
SO(10) GUT models.

(ii) The $\mu\to e+\gamma$ appears to be the only other low energy
test of the model which is similar to such models.


(iii) For the choice of parameters used in fermion mass fitting
the neutrino mixing angles and mass differences could have any
values in the allowed region.

\section{Conclusion}
In conclusion, we have presented the minimal SO(10) {\bf 16}-Higgs
model for fermion masses where the problem of extreme fine tuning of
higher dimensional Planck scale induced proton decay operators has been
considerably ameliorated by the presence of discrete symmetries so that
in the end, we only need to tune down the coupling only
by a factor of $10^{-2}$. In this sense it is a more natural model
We exhibited a fit to all fermion masses and mixings including
neutrinos in this model to show that it can indeed be a realistic
description of nature.

 This work is supported by the National Science Foundation
Grant No. PHY-0652363. This work was presented at the GUT2007
workshop at Ritsumeikan University in Japan by R. N. M. and we
like to thank the participants at this workshop and M. Ratz for
comments.




\begin{thebibliography}{90}

\bibitem{126}  T.~E.~Clark, T.~K.~Kuo and N.~Nakagawa,
  Phys.\ Lett.\  B {\bf 115}, 26 (1982);
 C. S. Aulakh and R. N. Mohapatra, Phys. Rev. {\bf
D 28}, 217 (1983);   K. S. Babu and R. N. Mohapatra, Phys. Rev.
Lett. {\bf 70}, 2845 (1993);
 D. G. Lee and R. N. Mohapatra, Phys. Rev. {\bf D 51},
1353 (1995); M.~C.~Chen and K.~T.~Mahanthappa, Phys.\ Rev.\  D
{\bf 62}, 113007 (2000); T. Fukuyama and N. Okada, hep-ph/0206118;
B. Bajc, G. Senjanovi\'c and F. Vissani, Phys.\ Rev.\ Lett.\ {\bf
90}, 051802 (2003) hep-ph/0210207. H.~S.~Goh, R.~N.~Mohapatra and
S.~P.~Ng, Phys.\ Lett.\ B {\bf 570}, 215 (2003)
[arXiv:hep-ph/0303055]; T.~Fukuyama, A.~Ilakovac, T.~Kikuchi,
S.~Meljanac and N.~Okada,
  JHEP {\bf 0409}, 052 (2004);Eur.Phys.J.C42:191-203,2005;
 B. Dutta, Y. Mimura and R. N. Mohapatra,
hep-ph/0406262, Phys. Lett. {\bf B 603 }, 35 (2004) ; Phys. Rev.
Lett. {\bf 94}, 091804 (2005); Phys. Rev. {\bf D 72}, 075009
(2005); B. Bajc, A. Melfo, G. Senjanovic and F. Vissani, Phys.
Lett. {\bf B 634}, 272 (2006); C. S. Aulakh and S. K. Garg,
hep-ph/0512224;  K. S. Babu and C. Macesanu, Phys. Rev. {\bf D
72}, 115003 (2005); S. Bertolini, M. Malinsky and T. Schwetz,
Phys. Rev. {\bf D 73}, 115012 (2006);
  W. Grimus and H. Kuhbock, hep-ph/0612132; arXiv:0710.1585 [hep-ph].

\bibitem{so1016} S.~M.~Barr and S.~Raby, Phys.\ Rev.\ Lett.\  {\bf 79},
4748 (1997) C. Albright, K. S. Babu and S. Barr, Phys. Rev. lett.
{\bf 81}, 1167 (1998); C. Albright and S. Barr, Phys. Rev. {\bf D
58}, 013002 (1998); K. S. Babu, J. C. Pati and F. Wilczek,
hep-ph/9812538; S.~Raby, Phys.\ Rev.\  D {\bf 65}, 115004 (2002);
R.~Dermisek and S.~Raby, Phys.\ Lett.\  B {\bf 622}, 327 (2005);
X. Ji, Y. Li and R. N. Mohapatra, Phys. Lett. {\bf B 633}, 755
(2006); R.~Dermisek, M.~Harada and S.~Raby, Phys.\ Rev.\ D {\bf
74}, 035011 (2006);S.~Morisi, M.~Picariello and E.~Torrente-Lujan,
Phys.\ Rev.\  D {\bf 75}, 075015 (2007)



\bibitem{A1} I.~Hinchliffe and T.~Kaeding, Phys. Rev. \textbf{D47}
(1993), 279; H.~K. Dreiner, C.~Luhn, and M.~Thormeier, Phys. Rev.
\textbf{D73} (2006), 075007, hep-ph/0512163; L.~E. Ib{\'a}{\~n}ez
and G.~G. Ross, Phys. Lett. \textbf{B260} (1991), 291;
 T.~Banks and M.~Dine, Phys. Rev. \textbf{D45} (1992), 1424,
hep-th/9109045;  K.~S.~Babu, I.~Gogoladze and K.~Wang,
  Phys. Lett.  \textbf{B570} (2003), 32, hep-ph/0306003;
  K.~Kurosawa, N.~Maru and T.~Yanagida, Phys. Lett.  \textbf{B512} (2001), 203,
  arXiv:hep-ph/0105136;  Y.~Kajiyama, E.~Itou and J.~Kubo,
  Nucl. Phys.  \textbf{B743} (2006), 74, hep-ph/0511268.

\bibitem{GUT} For other approaches to suppressing proton decay, see
P. Nath and R. Sayed,  Phys. Rev. \textbf{D 77} (2008) 015015; H.
Dreiner and M. Thormeir, Phys. Rev. \textbf{D 69} (2004) 053002;
J. Sayre, S. Wiesenfeldt and S. Willenbrock, Phys. Rev. \textbf{D
75} (2007) 037702; for a review of proton decay, see Pran Nath and
Pavel Fileviez Perez, Phys.Rept. \textbf{441} (2007)191-317.

\bibitem{rmmr}  R.~N.~Mohapatra and M.~Ratz,
  arXiv:0707.4070 [hep-ph], Phys.\ Rev.\  D {\bf 76}, 095003 (2007).

\bibitem{pdecay} K.~S.~Babu and S.~M.~Barr, Phys.\ Rev.\  D {\bf 48},
5354 (1993); Z.~Chacko and R.~N.~Mohapatra, Phys.\ Rev.\  D {\bf
59}, 011702 (1999); R.~Dermisek, A.~Mafi and S.~Raby,
  Phys.\ Rev.\  D {\bf 63}, 035001 (2001).



\bibitem{Fukuyama}
  T.~Fukuyama, A.~Ilakovac, T.~Kikuchi, S.~Meljanac and N.~Okada,
  J.\ Math.\ Phys.\  {\bf 46}, 033505 (2005)
  [arXiv:hep-ph/0405300].

\bibitem{Das}
  C.~R.~Das and M.~K.~Parida,
  Eur.\ Phys.\ J.\  C {\bf 20}, 121 (2001)
  [arXiv:hep-ph/0010004].


\bibitem{nfit}
  M.~Maltoni, T.~Schwetz, M.~A.~Tortola and J.~W.~F.~Valle,
  New J.\ Phys.\  {\bf 6}, 122 (2004)
  [arXiv:hep-ph/0405172].

\bibitem{Hisano}
    F.~Borzumati and A.~Masiero,
  Phys.\ Rev.\ Lett.\  {\bf 57}, 961 (1986).
  J.~Hisano, T.~Moroi, K.~Tobe, M.~Yamaguchi and T.~Yanagida,
  Phys.\ Lett.\  B {\bf 357}, 579 (1995)
  [arXiv:hep-ph/9501407];
 J.~Hisano, T.~Moroi, K.~Tobe and M.~Yamaguchi,
  Phys.\ Rev.\  D {\bf 53}, 2442 (1996)
  [arXiv:hep-ph/9510309].

\bibitem{masiero:2007ha}

 A.~Masiero, S.~K.~Vempati and O.~Vives,
  arXiv:0711.2903 [hep-ph];
  M.~Ciuchini, A.~Masiero, P.~Paradisi, L.~Silvestrini, S.~K.~Vempati and O.~Vives,
  Nucl.\ Phys.\  B {\bf 783}, 112 (2007)
  [arXiv:hep-ph/0702144].


\bibitem{meg} C. Bemporado et al. MEG Collaboration; PSI-R-99-05.

\end{thebibliography}
\end{document}